\documentclass[aps,reprint,nofootinbib]{revtex4-1}
\usepackage{graphicx,amsfonts,amsmath,amssymb}
\usepackage{bm}

\begin{document}
\title{On the ``Universality'' of the Form of Maxwell's Equations}
\date{\today}
\author{C. Baumgarten}
\affiliation{Switzerland}
\email{christian-baumgarten@gmx.net}

\def\begeq{\begin{equation}}
\def\endeq{\end{equation}}
\def\bquo{\begin{quotation}}
\def\equo{\end{quotation}}
\def\begary{\begeq\begin{array}}
\def\endary{\end{array}\endeq}
\newcommand{\myarray}[1]{\begin{equation}\begin{split}#1\end{split}\end{equation}}
\def\bmtx{\left(\begin{array}}
\def\emtx{\end{array}\right)}
\def\d{\partial}
\def\h{\eta}
\def\w{\omega}
\def\W{\Omega}
\def\s{\sigma}
\def\eps{\varepsilon}
\def\e{\epsilon}
\def\a{\alpha}
\def\b{\beta}
\def\g{\gamma}
\def\y{\gamma}
\def\d{\partial}
\def\S{{\Sigma}}

\def\leftD#1{\overset{\leftarrow}{#1}}
\def\rightD#1{\overset{\rightarrow}{#1}}

\begin{abstract}
  Many papers have been published over the years that either conjecture
  or even (claim to) prove the universality of the form of Maxwell's equations.
  We present yet another derivation of Maxwell's equations and discuss the
  conclusions suggested by Maxwell universality, namely the logical inevitability
  of the Lorentz transformations and the mathematical inconsistency of Newtonian
  physics.
\end{abstract}
\maketitle

%%%%%%%%%%%%%%%%%%%%%%%%%%%%%%%%%%%%%%%%%%%%%%%%%%%%%%%%%%%%%%%%%%%%%%%%%%%%%%%%
\section{Introduction}
%%%%%%%%%%%%%%%%%%%%%%%%%%%%%%%%%%%%%%%%%%%%%%%%%%%%%%%%%%%%%%%%%%%%%%%%%%%%%%%%

Some years ago, Burns provided yet another proof that the form of Maxwell's equations
is ``universal''~\cite{Burns2019} in the sense that Maxwell's equations can be
mathematically derived from the continuity equation and the dimension of
space-time. This means that whenever a locally conserved (``substantial'') quantity
like charge or mass moves in a space with three spatial (and one temporal)
dimension(s), it will be the source of vector fields of the Maxwellian form.

The question whether Maxwell's equation have one of many possible
forms of linear coupled partial differential equations, or whether their
form can be derived, has been subject in numerous publications in the past.
The first paper on this subject, known to the author, was written by H. Hertz
and dates back to 1884 (!)~\cite{Hertz1884}. But Hertz did {\it not} claim to
have proven the unique and universal form of Maxwell's equations (see also
Refs.~\cite{Zatzkis1965,Havas1966}).
Many more papers on the subject followed since then, for instance by Fritz Bopp
(1962)~\cite{Bopp1962}, Elliott Krefetz (1970)~\cite{Krefetz},
Jack Cohn (1978)~\cite{Cohn1978}, several papers by
D.H. Kobe~\cite{Kobe1978,Kobe1980,Kobe1986}, by
E. Kapuscik (1985)~\cite{Kapuscik1985},
Feynman (published by Dyson 1990)~\cite{Dyson1990},
Hehl et al. \cite{hehl1999}, J.S. Marsh~\cite{Marsh1993},
R.D. Nevels (1995)~\cite{Nevels1995}, A.M. Davis (2006)~\cite{Davis2006},
Jose A. Heras (2007,2009)~\cite{Heras1,Heras2},
B.P. Kosyakov (2014)~\cite{Kosyakov2014}, C. Baumgarten (2017)~\cite{qed_paper,osc_paper},
D.H. Sattinger (2018)~\cite{Sattinger2018}, L. Burns (2019)~\cite{Burns2019},
and J. K\"onig (2021)~\cite{Koenig2021}. This list is probably still incomplete.

Regarding the number of publications, the idea of mathematical universality
is not alien to the worldview of physics. It has been conjectured since long,
possibly since Helmholtz's theorem is known. 
Not all of the mentioned authors claimed to have given a rigorous derivation
of Maxwell's equations and not all authors used the same initial assumptions.
Insofar the issue might not yet be closed.

In Sec.~\ref{sec_MEQ} we shall provide a heuristic derivation
of Maxwell's equations starting from Bopp's version of 1962.
In Sec.~\ref{eq_GTR} we shall compare the electromagnetic Maxwell
equations with those resulting from the linear approximation of
general relativity as derived in Gron and Hervik~\cite{GronHervik}.
Besides the fact that general relativity has, in linear approximation,
Maxwell's form, also the reverse account seems possible, namely to
interpret electromagnetism as ``spacetime pseudo-curvature''~\cite{Bouhrik2025}.
In Sec.~\ref{sec_summary} we shall draw some preliminary conclusions.

\section{A Heuristic Derivation of Maxwell's Equations}
\label{sec_MEQ}

We start with the same arguments as provided by Fritz Bopp
in 1962~\cite{Bopp1962}. We assume a normalizable
scalar density $\rho(\vec x,t)$ for which the equation of
continuity holds
\begeq
{\d\rho\over\d t}+\vec\nabla\cdot\vec j=0
\label{eq_continuity}
\endeq
where $\vec j=\rho\,\vec v$ is the current density which
belongs to $\rho$. What the continuity equation {\it says}, is
little more than the mathematical definition for a material substance:
If the amount of substance in some volume changes, then there must
be a corresponding net material flow through the surface of the volume.
The assumption expressed by the continuity equation is not speculative
or arbitrary; it is inevitably required in any world that we would
accept as ``physically real''.

The density can always be expressed as the divergence of a
field $\vec E$, i.e. by $\rho=\eps_0\,\vec\nabla\cdot\vec E$.
Inserting this into Eq.~\ref{eq_continuity} gives:
\begeq
\eps_0\,{\d(\vec\nabla\cdot\vec E)\over\d t}+\vec\nabla\cdot\vec j=0
\endeq
so that
\begeq
\vec\nabla\cdot(\eps_0\,{\d\vec E\over\d t}+\vec j)=0\,.
\endeq
Then it follows from Helmholtz' theorem, that
the expression $\eps_0\,{\d\vec E\over\d t}+\vec j$
must be expressable by the rotation of some vector field
$\vec B$, so that one readily obtains Maxwell's second
inhomogeneous equation:
\begeq
\vec\nabla\times\vec B=\mu_0\,(\eps_0\,{\d\vec E\over\d t}+\vec j)
\endeq
The reasoning that leads to the homogenous equations, is slightly
more involved. But a symmetry analysis of the skalar and vector fields
introduced so far, provides some conditions for the possible
form of these equations.

\subsection{Time-Reversal and Parity-Symmetries of Scalars and Vectors}

The conventional vector calculus of Gibbs and Heaviside does not
generate or include information about the symmetry properties of
the physical quantities that are represented by these vectors~\footnote{
  This is different in Clifford algebraic representations of
  space-time, which provide ``vectors'' of different type and
  transformation behavior~\cite{lt_paper}.}.
Using this calculus the analyze of the symmetries of scalars
and vectors has to be done ``by hand'' on the basis of mathematical and/or
physical arguments. The density of matter (or charge) $\rho$, for instance,
may neither depend on the orientation of the coordinates-system (i.e. on parity)
nor on the chosen direction of time. Other quantities, like a velocity
vector or a current flow, must reverse sign on time reversal {\it and}
parity transformation.

In order to formalize the analysis of these symmetries, one can
introduce a time-reversal operator ${\cal T}$. The following properties
\myarray{
{\cal T}(t)&=-t\\
{\cal T}(\d t)&=-\d t\\
{\cal T}(\vec r)&=\vec r\\
{\cal T}(\rho)&=\rho\\
}
are evident. Accordingly one defines a parity operator ${\cal P}$:
\myarray{
{\cal P}(t)&=t\\
{\cal P}(\vec r)&=-\vec r\\
{\cal P}(\vec\nabla)&=-\vec\nabla\\
{\cal P}(\rho)&=\rho\\
}
Both operators obey the following product rule
\myarray{
{\cal T}(a\,b)&={\cal T}(a)\,{\cal T}(b)\\
{\cal P}(a\,b)&={\cal P}(a)\,{\cal P}(b)\\
}
Consequently, the velocity $\vec v$ is skew-symmetric with
respect to both, parity and time reversal:
\myarray{
{\cal T}(\vec v)&=-\vec v\\
{\cal P}(\vec v)&=-\vec v\\
}
and the same holds for the current density $\vec j=\rho\,\vec v$
since $\rho$ is invariant under both, time-reversal and parity.
Note that we do not intend here to make claims about a real
{\it physical} time-reversal. These operators concern the
{\it description} of the {\it same} physical situation with respect
to our free choice of the orientation of the used coordinate
system and time axis.

Nonetheless, when terms are added in equations, the parity and
time-reversal symmetry properties of all terms should be identical.
For instance, if one defines a vector $\vec E$ by
$\rho=\eps_0\,\vec\nabla\cdot\vec E$, the field $\vec E$ must
have an eigenvalue of $+1$ under time reversal and a parity of
$-1$.
\begin{table}[h]
\begin{tabular}{|l|l|c|c|}\hline
Scalar         & Vector                   & Parity & Time-Reversal \\\hline
$\rho$, ${\vec\nabla\cdot\vec E}$ & ${\d\vec B\over\d t}$, ${\vec\nabla\times\vec E}$ & +1     & +1\\
$t$, $\d_t$    & $\vec B$                        & +1     & -1\\
               & $\vec r$, $\vec\nabla$, $\vec E$  & -1     & +1\\
${\vec\nabla\cdot\vec B}$ & $\vec v$, $\vec J$, ${\d\vec E\over\d t}$, ${\vec\nabla\times\vec B}$  & -1     & -1\\\hline
\end{tabular}
\caption[]{Time reversal and parity symmetry properties of 
various scalars and vectors.
\label{tab_vecsym}
}
\end{table}
Accordingly, from 
\begeq
\vec\nabla\times\vec B=\mu_0\,\vec J+\mu_0\,\eps_0\,{\d\vec E\over\d t}
\label{eq_ampere_maxwell}
\endeq
one derives the symmetry of ${\vec\nabla\times\vec B}$ and of
$\vec B$, ${\d\vec B\over\d t}$ and ${\vec\nabla\cdot\vec B}$.
The same applies to the derivatives of $\vec E$.
In this way we obtain the symmetry properties of all
scalar and vector quantities as shown in Tab.~\ref{tab_vecsym}.

The result is that ${\vec\nabla\cdot\vec B}$ is a scalar with
parity $-1$, called {\it pseudoscalar} and it is the only
{\it pseudoscalar} quantity to be found in Tab.~\ref{tab_vecsym}.
Hence it should be zero. It has been suggested that the divergence
of $\vec B$ might be equal to a ``magnetic charge density''
in analogy to the electric field, but we can not possibly
imagine a legitimate reason to postulate a fundamental pseudo-scalar
quantity which changes sign when the coordinate system is reversed.
And of course, possibly for this reason, no magnetic charge has ever
been found~\footnote{Note that within the theory of the Hamiltonian
  Dirac-Clifford algebra, pseudoscalars turn out to be
  skew-Hamiltonian~\cite{qed_paper}.}.
But even if someone would not understand or accept this argument,
it would still be legitimate to set ${\vec\nabla\cdot\vec B}=0$
as a kind of gauge, since $\vec B$ was introduced as a purely
rotational field and a magnetic charge density was neither presumed
nor is it required for the derivation. 

There are two more remaining quantities, namely
${\d\vec B\over\d t}$ and ${\vec\nabla\times\vec E}$.
If we do not wish to invent ad-hoc some new quantity, then
the only possible Ansatz for the remaining equation is
\begeq
{\d\vec B\over\d t}=\pm{\vec\nabla\times\vec E}
\endeq
Combining this with the other equations, we obtain:
\myarray{
 \pm\vec\nabla\times{\d\vec B\over\d t}&=\pm\mu\,\eps{\d^2\vec E\over\d t^2}\\
 \vec\nabla\times\vec\nabla\times\vec E&=\pm\mu\,\eps{\d^2\vec E\over\d t^2}\\
 -\vec\nabla^2 E+\vec\nabla(\vec\nabla\cdot\vec E)&=\pm\,\mu\,\eps{\d^2\vec E\over\d t^2}\\
 -\vec\nabla^2 E&=\pm\,\mu\,\eps{\d^2\vec E\over\d t^2}\\
}
Therefore the resulting equation for the field $\vec E$ is:
\begeq
 \pm\,\mu\,\eps{\d^2\vec E\over\d t^2}+\vec\nabla^2 E=0
 \endeq
 The type of equation now depends on the chosen sign in front and
 and the sign of the constants $\mu$ and $\eps$.
 Provided both constants are positive~\footnote{
   In Sec.~\ref{eq_GTR} it will turn out that both constants are
   negative in case of gravitation.}, then the upper sign leads to
 solutions that can not be normalized and are hence unphysical.
 But the lower sign leads to a system of coupled wave equations
such that $c=1/\sqrt{\mu\,\eps}$ is the phase velocity of these
waves. Thus we arrive (for positive values of $\mu$ and $\eps$)
at Faraday's law of induction:
\begeq
\vec\nabla\times\vec E=-{\d\vec B\over\d t}\,.
\label{eq_faraday}
\endeq
Since the field $\vec B$ can assumed to be divergence-free, one
can write it as follows
\begeq
\vec B=\vec\nabla\times\vec A
\endeq
where the field $\vec A$ is called ``vector-potential''.
Together with Faraday's law it follows that
\begeq
\vec\nabla\times\vec E=-\vec\nabla\times{\d\vec A\over\d t}\,.
\endeq
so that $\vec E$ and ${\d\vec A\over\d t}$ may only differ by
a gradient field:
\begeq
\vec E=-{\d\vec A\over\d t}-\vec\nabla\phi\,.
\endeq
Then it follows from Gauss' law that
\begeq
 \vec\nabla\cdot\vec E=-\vec\nabla\cdot{\d\vec A\over\d t}-\vec\nabla^2\phi=\rho/\eps\,.
\endeq
Since the divergence of the vector potential is still undefined,
we have the freedom to use the Lorentz gauge:
\begeq
\vec\nabla\cdot\vec A+\mu\,\eps\,{\d\phi\over\d t}=0
\endeq
so that one obtains decoupled wave equations for the potentials
$\phi$ und $\vec A$:
\myarray{
  (\frac{1}{c^2}\,{\d^2\over\d t^2}-\vec\nabla^2)\,\phi&={\rho\over\eps}\\
  (\frac{1}{c^2}\,{\d^2\over\d t^2}-\vec\nabla^2)\,\vec A&=\mu\,\vec j\\
}
Many textbooks introduce Maxwell's equations as ``postulates'' or ``axioms'',
and some even claim that this is the only possible and legitimate method.
Even though there is no doubt that any physical theory requires
experimental confirmation, purely mathematical considerations
provide a degree of certainty and a depth of confidence that
experimental results alone can not provide.

Einstein once wrote that~\cite{EinsteinBio}:
\bquo
A theory can be tested by experience, but there is no way from experience
to the construction of a theory. Equations of such complexity as are the
equations of the gravitational field can be found only through the discovery
of a logically simple mathematical condition that determines the equations
completely or almost completely.
\equo
The assumption of local continuity is such a condition and allows to derive
Maxwell's equation ``completely or almost completely''. Hence the form of
Maxwell's equations is universal, i.e. must hold in any $3+1$-dimensional
world in which the continuity equation holds.

\section{The Gravitational Maxwell's Equations}
\label{eq_GTR}

Even though Maxwell's theory is defined on a flat space-time, while
general relativity is based on Riemannian geometry, nonetheless the
linear approximation of general relativity should yield the same
Maxwellian form of equations since in the previous section we made no
explicit reference to electromagnetism. Hence mass density and mass flow
should generate, in first order, a gravito-magnetic field that is described
by a set of coupled equations of Maxwell's form.
And this is indeed the case~\cite{GronHervik}:
\myarray{
  {\bf E}_g&=-{\bf\nabla} \phi_g-{\d{\bf\tilde A}_g\over\d t}\\
  {\bf\tilde B}_g&={\bf\nabla}\times{\bf\tilde A}_g\\
  {\bf\nabla}\cdot{\bf E}_g&=-4\,\pi\,G\,\rho\\
  {\bf\nabla}\cdot{\bf\tilde B}_g&=0\\
  {\bf\nabla}\times{\bf E}_g&=-{\d{\bf\tilde B}_g\over\d t}\\
  {\bf\nabla}\times{\bf\tilde B}_g&=-\frac{4\,\pi\,G}{c^2}\,{\bf
    j}+\frac{1}{c^2}{\d{\bf E}_g\over\d t}\\
  \label{eq_gm}
}
In order to demonstrate the exact formal equivalence with Maxwell's equations,
we used $\frac{1}{2\,c}{\bf A}\to{\bf\tilde A}$ and $\frac{1}{2\,c}{\bf
  B}\to{\bf\tilde B}$ compared ref.~\cite{GronHervik}.
If one then replaces $-\frac{4\,\pi\,G}{c^2}\equiv\mu_g\to\mu_0$ and
$-4\,\pi\,G\equiv 1/\eps_g\to 1/\eps_0$, then
one obtains the standard form of Maxwell's equations with $1/c^2=\mu_0\,\eps_0$. 
%\myarray{
 % {\bf E}&=-{\bf\nabla} \phi-{\d{\bf A}\over\d t}\\
  %{\bf B}&={\bf\nabla}\times{\bf A}\\
  %{\bf\nabla}\cdot{\bf E}&=\rho/\eps_0\\
  %{\bf\nabla}\cdot{\bf B}&=0\\
  %{\bf\nabla}\times{\bf E}&=-{\d{\bf B}\over\d t}\\
%{\bf\nabla}\times{\bf B}&=\mu_0\,{\bf j}+\frac{1}{c^2}{\d{\bf E}\over\d t}\\
%}
The SI unit of $E_g$ is that of acceleration $\rm{m/s^2}$, the unit of
${\tilde B}_g$ is $\rm{s^{-1}}$, i.e. it has the unit of frequency.
The gravitational Lorentz force is then given by
\begeq
{\bf F}=m\,({\bf E}_g+{\bf v}\times{\bf\tilde B}_g)\,,
\endeq
so that for $v\ll c$ one has
\begeq
{\bf a}={\bf E}_g+{\bf v}\times{\bf\tilde B}_g\,.
\endeq
The gravitational analog of the electric field is hence a vector field
of pure acceleration and the gravitational analog of the magnetic field
is a field of pure rotation.

Einstein developed the theory of general relativity on the basis of Riemannian
geometry. It is widely accepted today. This theory is highly non-linear and
hence one has to use the linear approximation in order to compare it with
Maxwell's theory. As it turns out, the linear approximation
of the gravitational field equations has the form of Maxwell equations, i.e.
contains gravito-magnetic field terms. Even though this is textbook
wisdom~\cite{GronHervik,Ryder2009,Straumann}, the
significance of this math fact is rarely ever emphasized. It would be a
very strange coincidence to find the non-trivial form of Maxwell's equations
in two different physical theories just merely by chance!

\section{Consequences of Maxwell-Universality}
\label{sec_summary}

Of course the proof that the form of Maxwell's equations is universal in 
$3+1$ dimensions does, as such, not provide certainty about the phenomena
of electromagnetism or gravity.
From the form alone, it is not possible to ``derive'' the existence of
positive and negative charges. Neither does it imply that we could trust
a mathematically derived possibility without experimental verification.
Nonetheless the univerality of the form of Maxwell's equations is
of paramount importance for the proper understanding of electromagnetism,
gravity and the meaning of the Lorentz transformations.
If the mere condition of continuity restricts the form of possible linear
field equations to Maxwell's form, then this provides a certainty regarding
the form of electrodynamics that comes close to finality.

Is it possible and/or acceptable that the confidence in the truth of physical
laws stems not exclusively from experimental verification but also
on the universality of certain mathematical forms? And if so, how
do we weight this degree of certainty?

A position of indifference in this question might be tenable from
an engineering perspective on physics: The theoretical inevitability of
the form of Maxwell's equations has little practical consequences for
the work of an experimental physicist or of an electrical engineer.
Nonetheless it substantially changes our understanding of these equations
and it should change the way in which we communicate physical laws to
students and/or to the interested public.
If it is true that certain forms of physical laws could not possibly be
different from what they actually are (unless space-time had a different
dimensionality or matter would not be conserved), then the curriculum
should clearly point this out. 

Newton declared in his {\it principiae} that the notions
of space, time and motion do not require definitions. They are given
as self-evident notions of common sense~\cite{NewtonPrincipiae}:
\begin{quotation}
I do not define time, space, place and motion, as being well known to all.
\end{quotation}
But since Maxwell's equations apparently can be derived from the
dimensionality of space-time plus continuity, then the common-sense
notion of absolute space and time is not just factually wrong,
but it is wrong for {\it mathematical} reasons. That is, Newton's
absolute space and time are mathematically inconsistent:
A Newtonian world is impossible even {\it in principle}.

Of course the laws of mechanics require {\it some} kind of relativity.
Loewdin has shown that, under very few reasonable assumptions, there
are only two possible types of transformations between inertial
frames, namely the Galilean and the Lorentz transformations~\cite{Loewdin98}.
Furthermore it has been demonstrated by Le Bellac and Levy-Leblond
that Maxwell's equation are incompatible with Galilean relativity~\cite{LeBellac1973}.
To be more precise, they have shown that the following three assumptions
are inconsistent:
\begin{enumerate}
\item Galilean invariance;
\item validity of continuity equation;
\item existence of magnetic forces between electric currents.
\end{enumerate}
The possibility to derive Maxwell's equations from continuity seems to
provide evidence that already the first two of these assumptions are
eventually inconsistent, though they seem to be compatible when
viewed in isolation.

To acknowledge Maxwell universality requires a scientific attitude
that is willing to accept a foundational role of mathematics in physics.
This point is problematic insofar as it concerns the falsifiability of
theories. If the mathematical form of a theory turns out to be inevitable,
what does this imply for it's falsifiability? What does it imply with respect
to the question of finality in physics? Is it scientifically legitimate to
simply ignore obvious math facts in order to preserve the position
that also a mathematically inevitable theory must be regarded as falsifiable? 

It appears that the form of (at least some) laws of physics
can be derived from pure math facts combined with a reality constraint.
If a theoretical framework has the most general form, thus exhausting the
mathematical possibilities, then it generates a different degree of scientific
certainty than a framework that is merely based on experimental facts: it is
the difference between just knowing that something {\it is} the case and
knowing {\it why} something {\it should be} the case. It is also clear that
even the latter is not enough: Even if one knows that and why something
{\it should be} the case, one will always want to test whether it {\it really}
is the case. This is the reason why experimental confirmation is still required:
We can not fully trust theory alone, since inconsistencies can be hidden in
seemingly harmless assumptions~\footnote{Also classical electromagnetism has
  to fight with inconsistencies if based on point particles~\cite{Lechner}.
These inconsistencies are due to apparent infinities and are thus
very different from the inconsistency of the common-sense ideas
of space, time and matter.}.

Historically Maxwell's equations were found with the help of various
experiments. But this does not imply that experiments are specifically useful
to understand the rigidity of and the deeper logic behind their mathematical
form and symmetry. Many textbooks provide only a limited perspective on the
logical connection between continuity and Maxwell's equations: Maxwell's
equations are simply postulated and the continuity equation is derived
from them in passing. But maybe this perspective is too narrow and requires
revision.

\bibliography{meq_paper.bib}{}
\bibliographystyle{unsrt}

\end{document}